\documentclass[preprint,showpacs,preprintnumbers,amsmath,amssymb]{revtex4}
\usepackage{dcolumn}
\usepackage{bm}
\usepackage{epsfig}
\begin{document}

\hfill {\bf November 2005}

\title{Image method solutions for free-particle wave packets 
\\
\vskip 0.05cm
in wedge geometries}

\author{R. W. Robinett} \email{rick@phys.psu.edu}
\affiliation{%
Department of Physics\\
The Pennsylvania State University\\
University Park, PA 16802 USA \\
}

\date{\today}

\begin{abstract}
Using analogs of familiar image methods in electrostatics and optics, 
we show how to construct closed form wave packet solutions of the 
two-dimensional free-particle Schr\"{o}dinger equation in geometries 
restricted by two infinite wall barriers separated by an angle 
$\Theta_{N} = \pi/N$. As an example, we evaluate probability densities
and expectation values for a zero-momentum wave packet solution 
initially localized in a $\Theta_{3} = \pi/3 = 60^{\circ}$ wedge. 
We review the time-development of zero-momentum wave packets placed near 
a single infinite wall barrier in an Appendix.
\end{abstract}

\maketitle

\section{Introduction}
\label{sec:introduction}

Examples of closed-form wave packet solutions of the time-dependent 
Schr\"{o}dinger equation (SE) in introductory quantum mechanics are 
at a premium. A few standard books treat the case of wave packets for
the harmonic oscillator (either coherent or squeezed state versions),
often using methods which are rather advanced, such as propagator 
techniques \cite{saxon}. 
Almost all standard texts, however, develop the explicit example of 
the free-particle Gaussian wave packet, which requires only simple 
integration, and which yields a closed form solution. These solutions 
can, in turn, be used to easily evaluate many expectation values and 
examine both semi-classical propagation as well as wave packet spreading.

A simple generalization of that example to a slightly
more `interacting' case is that of the single infinite wall in one-dimension 
(or particle on the half-line) defined by the potential
\begin{equation}
V(x) = 
\left\{
\begin{array}{cc}
   0   & \qquad x > 0  \\
\infty & \qquad x \leq 0
\end{array}
\right.
\, . 
\end{equation}
Andrews \cite{andrews} has noted that simple linear combination
solutions of the form
\begin{equation}
\tilde{\psi}(x,t) = 
\left\{
\begin{array}{ll}
\psi(x,t) - \psi(-x,t) & \quad x \geq 0  \\
0                      & \quad x \leq 0
\end{array}
\right.
\label{original_mirror_solution}
\end{equation}
will also satisfy the Schr\"{o}dinger equation for $x>0$, provided 
that $\psi(x,t)$ is a 1D free-particle solution, since the free-particle
Hamiltonian is invariant under $x \longleftrightarrow -x$. This form is
then seen to also satisfy the necessary boundary condition at the 
infinite wall barrier for all times.  Such mirror solutions have been 
used to pedagogically examine the detailed nature
of the `collision' of a wave packet with such an infinite wall barrier
\cite{doncheski_1}, for which many exact results are possible
\cite{doncheski_2}, as well as motivating discussions of possible 
experimental tests of the problem of the 
``{\it Deflection of quantum particles by impenetrable boundary}'' 
\cite{dodonov}.

A simple generalization of this `mirror' problem to two-dimensions involves
two infinite wall barriers placed at right angles ($90^{\circ}$), 
defined by a potential 
\begin{equation}
V_{90} = 
\left\{
\begin{array}{ll}
   0   & \quad \mbox{for $x>0$ and $y>0$} \\
\infty & \quad \mbox{otherwise}
\end{array}
\right.
\label{corner_reflector}
\end{equation}
where the particle is constrained to be in the first (upper-right) quadrant.
Product solutions,  using `mirror' combinations as above, of the form
\begin{equation}
\tilde{\psi}(x,y;t) =
\left\{
\begin{array}{ll}
[\psi_{1}(x,t) - \psi_{1}(-x,t)]
[\psi_{2}(y,t) - \psi_{2}(-y,t)]
& \quad \mbox{for $x\geq 0$ and $y\geq 0$} \\
0 & \quad \mbox{otherwise}
\end{array}
\right. 
\end{equation}
satisfy the 2D Schr\"{o}dinger equation for $x,y>0$,  as well as the boundary
conditions on  the two intersecting infinite barriers. The form of the
product wavefunction in the first quadrant,
\begin{eqnarray}
\tilde{\psi}_{1}(x,t) \tilde{\psi}_{2}(y,t)
& = & 
\psi_{1}(x,t)\psi_{2}(y,t)
-\psi_{1}(x,t)\psi_{2}(-y,t) 
-\psi_{1}(-x,t)\psi_{2}(y,t) \\ \nonumber
& & 
\qquad
\qquad
\qquad
+\psi_{1}(-x,t)\psi_{2}(-y,t)
\end{eqnarray}
suggests that an even more general form, namely
\begin{equation}
\tilde{\psi}(x,y;t)
= 
  \xi(x,y;t)
-\xi(x,-y;t)
-\xi(-x,y;t)
+\xi(-x,-y;t)
\label{general_solution}
\, ,
\end{equation}
will satisfy the appropriate SE (for any $\xi(x,y;t)$ which does) 
since the 2D free-particle Hamiltonian is invariant under both 
$x \longleftrightarrow -x$ and $y \longleftrightarrow -y$. 
The additional terms 
in Eqn.~(\ref{general_solution}) correspond to two `mirror' terms 
reflected in each infinite wall barrier (flipped in sign) with an 
additional term (with the original sign) arising from two reflections, 
all terms being necessary in order to satisfy the boundary conditions 
on both infinite wall barriers.

Thought of in this way, such solutions are immediately reminiscent of
image solutions to the related problem of two perpendicular conducting plates
in electrostatics or the even more fundamental problem of the images
of a single real object as seen in a pair of mirrors placed at right angles.
The pattern of `signs' used in each of those two cases is identical to that
seen in Eqn.~(\ref{general_solution}). This suggests that a much wider
variety of `two-wall' problems in quantum mechanics can be constructed
using ones intuition from such classical textbook examples.

The potential in Eqn.~(\ref{corner_reflector}) is a specific case of two
infinite barriers intersecting with an arbitrary angle $\Theta$,  defined 
by
\begin{equation}
V_{\Theta} (r,\theta) = 
\left\{
\begin{array}{ll}
0 & \quad \mbox{if $r>0$ and $0<\theta<\Theta$} \\
\infty & \quad \mbox{otherwise}
\end{array}
\right.
\, . 
\end{equation}
The corresponding electrostatic (or optics) problems for the special cases
of $\Theta_{N} = \pi/N$ can be solved by making use of a finite number of
virtual `image' charges (or true images) given by the following prescription.
Place the original charge (or object) at a general point $(x,y)$ within the
allowed wedge region. Add a `mirror' image charge (or object) at $(x,-y)$,
but with opposite sign (handedness). 
Then repeatedly ($N-1$ times) rotate the location of these two charges 
(or objects) through $\theta = 2\Theta_{N}$, keeping track of the appropriate
sign (or handedness). For $\Theta = \Theta_{N} = \pi/N$, this process closes 
on itself, and the resulting set of $2N$ charges (or objects),
$2N-1$ of which are auxiliary ones, satisfies the
boundary conditions for the problem. (The cases of $N=1,2$ correspond to
the single infinite wall and the $90^{\circ}$ wedge in 
Eqn.~(\ref{corner_reflector}) respectively.)
In the context of the 2D Schr\"{o}dinger equation, such rotations of
coordinates can be easily proved to leave the free-particle Hamiltonian 
invariant, so that a wavefunction of the form 
$\psi(x\cos(\phi) + y\sin(\phi), -x\sin(\phi) + y \cos(\phi);t)$ will
be a solution if $\psi(x,y;t)$ is. This fact justifies the use of 
additional rotated versions of an existing free-particle solution in a
linear combination,  generalizing the result Eqn.~(\ref{general_solution}) 
for the case of $N=2$.

\begin{figure}
\epsfig{file=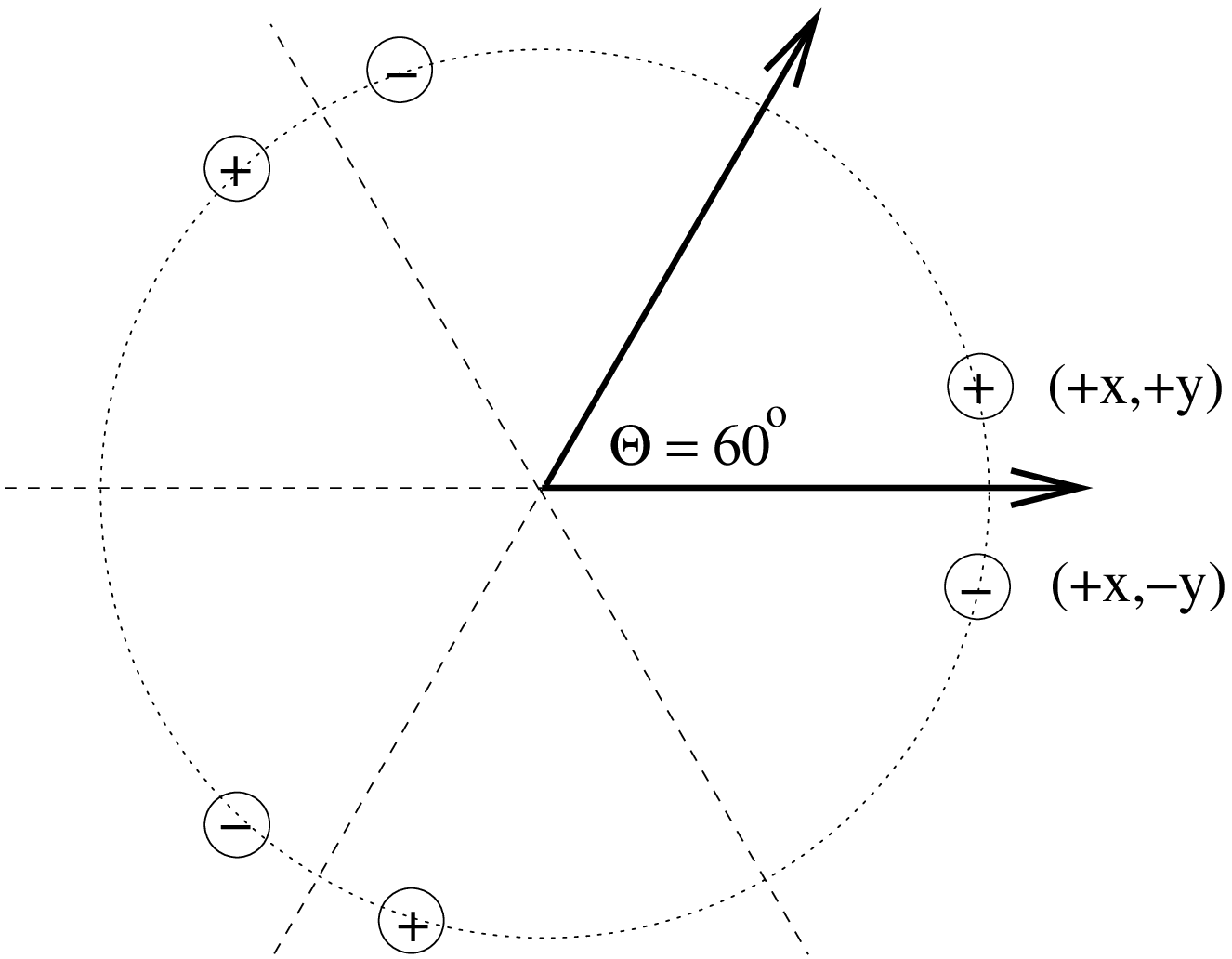,width=10cm,angle=00}
\caption{Image construction for the $\Theta_{3} = \pi/3 = 60^{\circ}$ 
wedge potential, leading to the solution in Eqn.~(\ref{exact_solution}).}
\label{fig:wedgepix}
\end{figure}

As an example of this construction and its relevance to the problem
of two infinite wall barriers, consider the case of $\Theta_{3} = 60^{\circ}$,
as shown in Fig.~\ref{fig:wedgepix},  where the prescription  above yields
the five 'image' points shown there. Using the appropriate locations, 
and signs, it is possible to construct a general solution as
\begin{eqnarray}
\psi_{60}(x,y;t) & = & 
\qquad
\qquad
\qquad
\psi(x,y;t) - \psi(x,-y;t) \nonumber \\ 
& &
+ \psi\left(-\frac{x}{2} - \frac{\sqrt{3}y}{2},\frac{\sqrt{3}x}{2} - \frac{y}{2};t\right) 
- \psi\left(-\frac{x}{2} + \frac{\sqrt{3}y}{2},\frac{\sqrt{3}x}{2} + \frac{y}{2};t\right) 
\nonumber \\
& &
+ \psi\left(-\frac{x}{2} + \frac{\sqrt{3}y}{2},-\frac{\sqrt{3}x}{2} - \frac{y}{2};t\right) 
- \psi\left(-\frac{x}{2} - \frac{\sqrt{3}y}{2},-\frac{\sqrt{3}x}{2} + \frac{y}{2};t\right) 
\label{exact_solution}
\,. 
\end{eqnarray}
This can readily be seen to satisfy the boundary conditions on the two 
infinite walls, namely
\begin{equation}
\psi_{60}(x,y=0;t) = 0 = \psi_{60}(x,y=\sqrt{3}x;t)
\qquad
\forall x,t
\, . 
\end{equation}
The results for the case of $\Theta = \pi/4 = 45^{\circ}$ and others
are obtained in exactly the same way. We note that the 1D infinite
square well has been discussed in terms of image methods
\cite{kleber}, \cite{born}, requiring an infinite number of virtual
terms, analogous to the optical case of two parallel mirrors. The
$N \rightarrow \infty$ limit of small angle wedges 
($\Theta_{N} \rightarrow 0$) is qualitatively
different as it does not lead to bound states.

The standard 1D Gaussian free-particle solution with arbitrary initial
central position and momentum ($x_0$ and $p_0$) is given by
\begin{equation}
\psi_{(G)}(x,t)  =   \frac{1}{\sqrt{\sqrt{\pi} \beta (1+it/t_0)}}
\,
e^{ip_0(x-x_0)/\hbar}
\, e^{-ip_0^2t/2m\hbar}
\,
e^{-(x-x_0-p_{0}t/m)^2/2\beta^2(1+it/t_0)}
\label{gaussian_form}
\end{equation}
where 
\begin{equation}
t_0 \equiv \frac{m \beta^2}{\hbar}
\, ,
\qquad
\Delta x_0 = \frac{\beta}{\sqrt{2}}
\, ,
\qquad
\mbox{and}
\qquad
\Delta p_0 = \frac{\hbar}{\beta \sqrt{2}}
\label{spreading_time}
\end{equation}
are the spreading time and initial spreads in position and momentum
(for the free 1D particle) respectively.

\begin{figure}
\epsfig{file=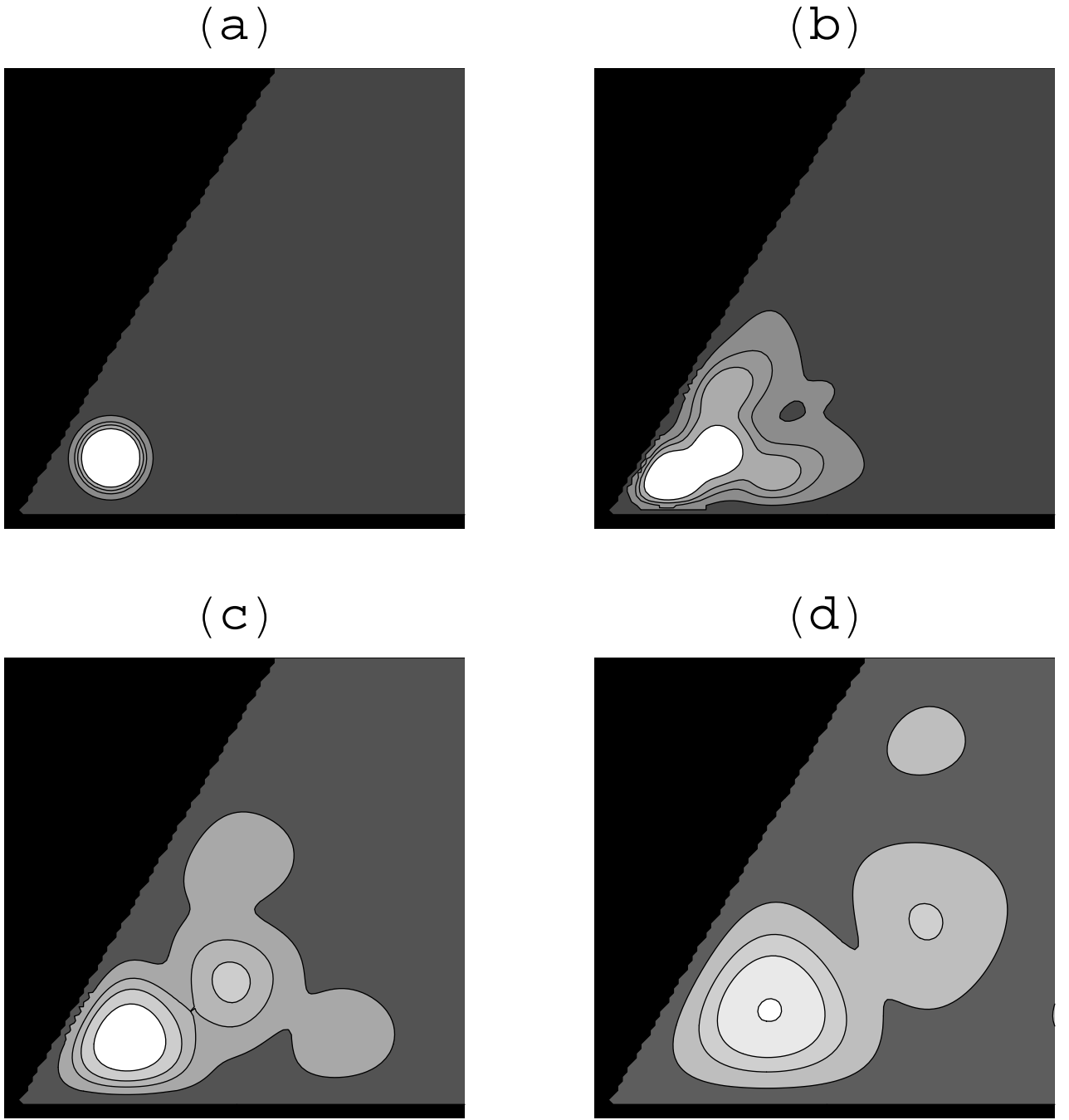,width=19cm,angle=00}
\caption{Contour plot of the probability density 
($|\psi(x,y;t)|^2$ versus $(x,y)$) for a spreading
wave packet in the $\Theta_{3} = 60^{\circ}$ wedge, 
using the solution in Eqn.~(\ref{exact_solution}) with individual terms
being a product of Gaussian forms as in Eqn.~(\ref{gaussian_form}).
The example shown here has $(p_{x0},p_{y0}) = (0,0)$ and
$(x_0,y_0) = (5,3)$. Numerical values of $\hbar,m,\beta = 1$ are 
used which give $t_0 = 1$ as well. The (a)-(d) cases shown correspond
to $t = 0,5,10,15$ in these units.}
\label{fig:53pix}
\end{figure}

Products of these of the form $\psi_{(G)}(x,t)\psi_{(G)}(y,t)$ can 
then be used in Eqn.~(\ref{exact_solution}) for any desired initial 
conditions. A simple example of a spreading wavepacket, corresponding to 
$(p_{y0}, p_{x0}) = (0,0)$, initially near the corner of a 
$\Theta_{3} = 60^{\circ}$ wedge,  is shown in Fig.~\ref{fig:53pix}, 
illustrating the complex time-development possible.

For the case of the single infinite wall, and the 
$\Theta = \pi/2 = 90^{\circ}$ wedge, such an approach can be continued 
to include the exact normalization of the wave function if Gaussian 
components are used as well as in the evaluation of a number of 
expectation values and related quantities exactly \cite{doncheski_2}. 
For the cases of $\Theta = \pi/N$ with $N > 2$, the
normalization integrals involve Airy functions and so cannot be done in
as compact a closed form, but experience with simpler cases guarantees
that if the individual `component' terms (the original and image 
contributions) are sufficiently separated in phase space,
i.e. if their central values of $x_0,p_0$ are far apart in units of the
fundamental position/momentum spreads, $\Delta x_{0}$ and $\Delta p_{0}$,
then the normalization factors differ from unity by exponentially small
terms. 

These systems can be profitably discussed in introductory quantum mechanics
courses, as non-trivial examples of the imposition of boundary conditions 
in a novel 2D geometry, in a free-particle context as opposed to more 
familiar bound state systems. 
In this context, the $\Theta = \pi/2$, $\pi/3$ and $\Theta = \pi/4$ cases can
be compared to bound state quantum wells with related geometries, as 
special cases of wave propagation near a `corner'. The $N=2$ case of
two walls at right angles provides an example of relevance to the 2D square well,
where the very short-term evolution of wave packets in that bound-state system
would be that of just such a `corner reflector', but with a more complex
long-term time-dependence, including the possibility of quantum revivals.
The $45^{\circ}-45^{\circ}-90^{\circ}$ triangular well (or
quantum billiard) is soluble using the odd linear combination of degenerate
eigenstates of the corresponding 2D square well, namely $\psi(x,y) =
(u_{n}(x)u_{m}(y) - u_{m}(x)u_{n}(y))/\sqrt{2}$ (where $m \neq n$) which
satisfies the boundary conditions on the sides of the square, but also along
the $y=x$ diagonal boundary. The case of an equilateral 
($60^{\circ}- 60^{\circ}-60^{\circ}$) triangle infinite well is also known
(perhaps not nearly as well as it should be) to also have simple closed
form energy-eigenstate solutions \cite{equilateral} and the 
$\Theta_{3} = 60^{\circ}$ wedge
can be thought of as one `corner' of that triangular billiard. 
(The construction of wave packets in all three bound state systems
has been discussed in a pedagogical context in Ref.~\cite{blueprint}.)

\begin{figure}
\epsfig{file=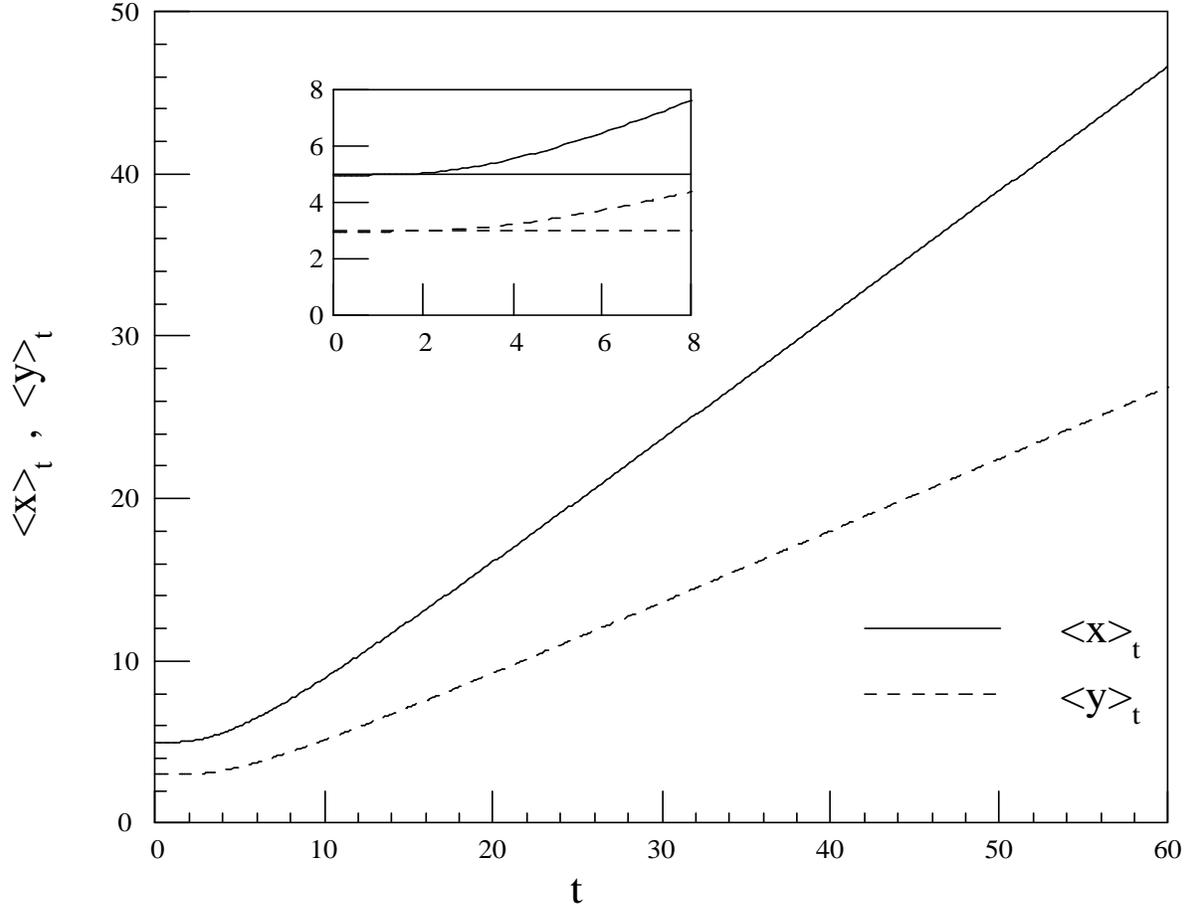,width=12cm,angle=270}
\caption{Expectation values, $\langle x \rangle_t$ and $\langle y \rangle_t$
versus $t$, for the zero-momentum initial state shown in 
Fig.~\ref{fig:53pix}. The insert shows the short-term time-development,
initially consistent with the (stationary) expansion of a zero-momentum 
wave packet, 
until the higher momentum components begin to reflect from the infinite 
wall boundaries, giving non-zero values of $\langle p_x \rangle_t$ and
$\langle p_y \rangle_t$.}
\label{fig:expectation_values}
\end{figure}

The ability to manipulate relatively simple closed-form solutions enables 
a wide variety of visualizations to be easily generated, so students can 
examine the analogs of many classical `bouncing' trajectories, including
more complex ones than the simple `in-and-out' paths for the $N=2$ `corner 
reflector'. More sophisticated projects involving the 
numerical evaluation of time-dependent expectation values (to be compared 
with classical expectations) or of the momentum-space probability densities
are now more within reach since a closed-form solution for $\psi(x,y;t)$ 
already exists. For example, while a classical particle placed near the
corner of a wedge would remain at rest, Fig.~\ref{fig:53pix} 
suggests that the long-term expectation values of $\langle x \rangle_t$ 
and $\langle p \rangle_t$ will have interesting non-classical behavior. 
To examine just this effect, the time-dependent expectation values 
$\langle x \rangle_t$ and $\langle y \rangle_t$ for that case are
evaluated numerically and shown in Fig.~\ref{fig:expectation_values} 
for the same parameter set as in Fig.~\ref{fig:53pix}. 
We note in this case that the long-term
values of $\langle p_x \rangle_t$ and $\langle p_y \rangle_t$, evaluated
using the basic relation $\langle p_x \rangle_t = m d\langle x \rangle_t/dt$,
can be seen to satisfy $\tan(\theta) = 
\langle p_y \rangle_t / \langle p_x \rangle_t \approx 0.577$ giving
$\theta = 30^{\circ}$ to a very high numerical accuracy; thus, the wave 
packet develops in time in a manner consistent with its restricted geometry. 
The interplay between the generation of non-zero values of the
expectation values of momenta and the conservation of kinetic energy
are questions which arise naturally to students when examining such systems.
These are perhaps more easily addressed in the context of a single infinite
wall (the $N=1$ case, as discussed in Ref.~\cite{dodonov}) and we review some
aspects of those questions in 
Appendix~\ref{appendix:momentum_and_kinetic_energy}.

Finally, and perhaps more importantly, the notion of multiple Gaussian 
wavepackets overlapping and demonstrating interference phenomena, such 
as seen in
Fig.~\ref{fig:53pix}, is a useful introduction or way to motivate the
study of famous experiments which have shown the 
``{\it Observation of interference between two Bose condensates}''
(as first discussed in Ref.~\cite{original_bec}, but also observed in 
Ref.~\cite{other_bec}, and subsequently extended to see matter wave 
interference between large numbers of BEC's \cite{bec_many}.)
 Such experiments have been analyzed
\cite{wallis} in terms of linear combinations of localized Gaussian wave
packets, allowed to expand after being released, and exhibiting the
observed interference behavior as they overlap. One can imagine 
\cite{robinett_bec} single 
BEC's released near one or more  infinite wall boundaries (modeling atomic 
mirrors) being described by a wave function solution much like that 
in Eqns.~(\ref{original_mirror_solution}), (\ref{general_solution}), or
(\ref{exact_solution}), discussed here in a more pedagogical context.

\appendix
\section{Momentum and kinetic energy for a zero-momentum 1D wave packet 
impinging on an infinite wall}
\label{appendix:momentum_and_kinetic_energy}

For a classical point particle rebounding elastically from a wall, the
impulsive change in momentum (from $\pm p_0$ to $\mp p_0$) and the fact 
that the total kinetic energy is unchanged (since $T = p^2/2m$), 
have  analogs in the interactions
of a quantum wave packet with a 1D infinite wall \cite{doncheski_1},
\cite{doncheski_2}. 
The time-development of the expectation values of momentum, kinetic energy,
and the spread in momentum for a initially `zero-momentum' wave packet
expanding near an infinite wall can easily raise related questions 
which can be nicely visualized, understood intuitively, and for which
long-time approximations can be derived, all aspects which we review in 
this Appendix.

\begin{figure}
\epsfig{file=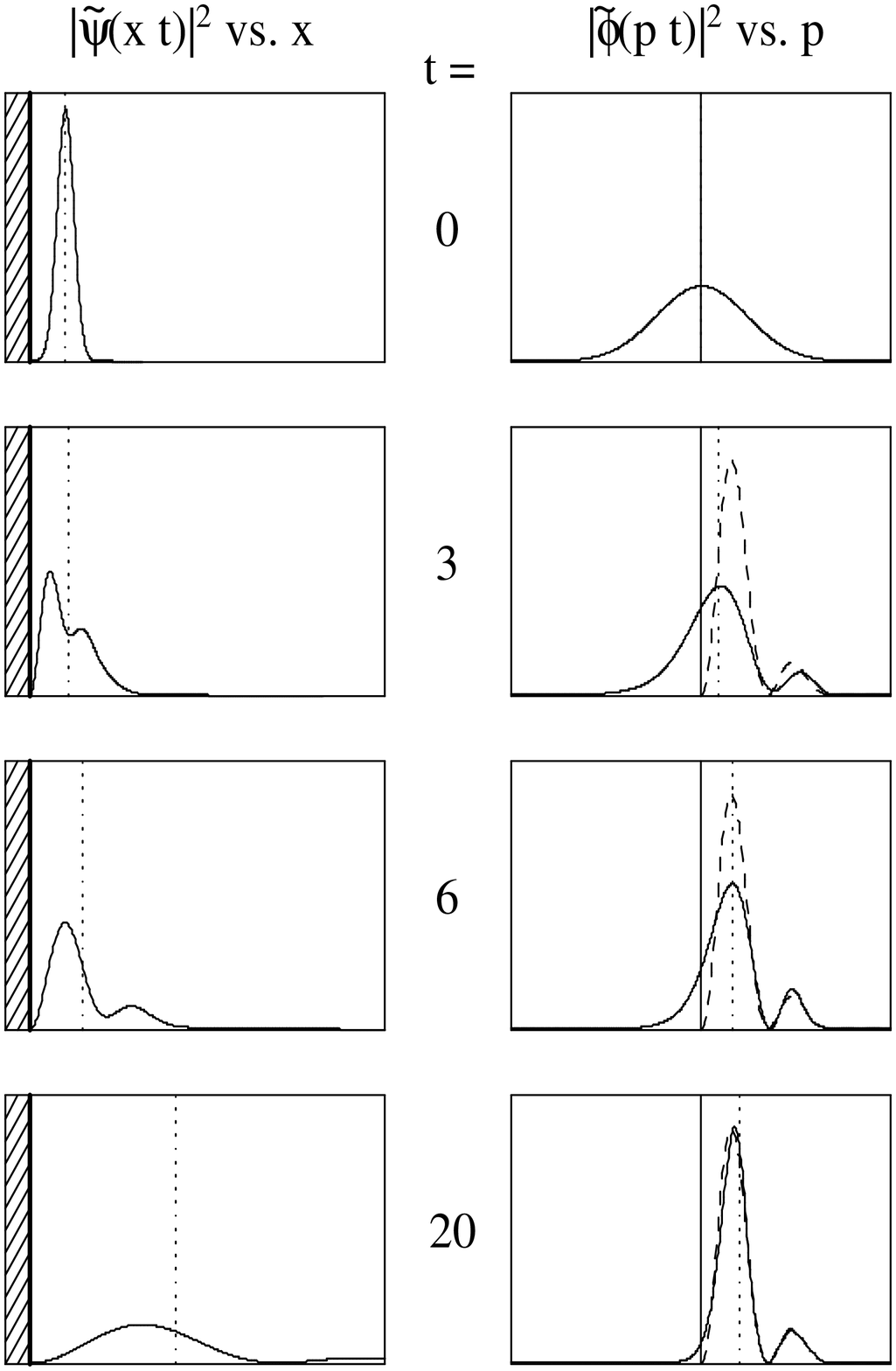,width=12cm,angle=000}
\caption{Plots of the position-space (left) and momentum-space (right)
probability densities versus time for a $p_{0x} = 0$ wavepacket allowed
to expand near an infinite wall. The vertical dotted lines show the 
time-dependent expectaion values $\langle x \rangle_t$ (left) and
$\langle p \rangle_t$ (right). The dashed curve on the right is the 
$t>>t_0$ solution from Eqn.~(\ref{long_term_solution}).
Values of $x_0 = 3$ and $\hbar,\beta,m = 1$ are used, giving $t_0 = 1$.}
\label{fig:zero_momentum}
\end{figure}

In Fig.~\ref{fig:zero_momentum}, we
plot the position-space ($|\tilde{\psi}(x,t)|^2$ versus $x$) and 
momentum-space
($|\tilde{\phi}(p,t)|^2$ versus $p$) distributions 
for a simple `mirror' solution of the
form in Eqn.~(\ref{original_mirror_solution}) using a Gaussian 
$\psi_{(G)}(x,t)$
with $p_0 = 0$. The momentum-space wave function is obtained by numerical
Fourier transform of the closed-form $\tilde{\psi}(x,t)$. A free-particle
$p_0=0$ Gaussian wave packet would simply spread, while maintaining 
its Gaussian form, with an increasing uncertainty in position given by 
$\Delta x_t = \Delta x_0 \sqrt{1 + (t/t_0)^2}$; the corresponding 
momentum-space probability distribution, $|\phi(p,t)|^2$, would remain 
unchanged, since $\phi(p,t) = \phi(p,0)\,\exp(-ip^2t/2m\hbar)$ for the 
1D free-particle.

For this case of the 1D infinite barrier, however, the wave packet components
which move towards the wall ($p<0$) must rebound, flipping sign and combine
with the right-moving ($p>0$) values which are not flipped, yielding an 
overall 
non-zero value of the expectation value of momentum, as shown by the 
vertical dotted lines in the $|\tilde{\phi}(p,t)|^2$ plots on the right
of Fig.~\ref{fig:zero_momentum}. 
The negative values of momentum in the initial $\tilde{\phi}(p,0)$ 
are, roughly speaking,  `folded over' on top of the positive values 
after reflection in sign, yielding a distribution which is therefore 
also narrower; a crude estimate based on this picture would be that the
momentum-space spread would be roughly half as large as its initial
value.  For increasingly 
large time, any small negative components will eventually rebound from 
the wall so that $|\tilde{\phi}(p,t)|^2$ becomes non-vanishing only 
for $p>0$.

For any free-particle solution, the expectation value of
the kinetic energy, given by $\langle T \rangle = \langle p^2\rangle/2m$,
will remain fixed, but the relation 
\begin{equation}
\frac{\hbar^2}{2\beta^2} \approx
\langle p^2 \rangle_0 =
\langle p^2 \rangle_t = \langle p \rangle_t^2 + (\Delta p_t)^2
\end{equation}
allows for a non-zero value of $\langle p\rangle_t$ to develop (as it
does here, for the physical reasons mentioned above) but at the expense 
of a smaller width in momentum-space (because the negative components
are `flipped' on top of the corresponding positive ones.)

The interference structure seen in the momentum-space probability density
can also be understood from simple arguments. Momentum components labeled
by $p$ will have an associated wavelength of $\lambda = 2\pi \hbar/p$ and
as the wave packet spreads (and interferes with its `mirror image') there
can be destructive interference between $\psi(x,t)$ and $\psi(-x,t)$ 
(destructive because of the sign difference in 
Eqn.~(\ref{original_mirror_solution})) 
when $n \lambda = 2x_0$ or when $p = n (\pi \hbar/x_0)$. (This type of
interference structure in the momentum-space probability density has
been studied in the context of Bose-Einstein condensates 
\cite{momentum_space} and is a general feature of the overlap of 
two (or more) expanding Gaussians.)

An excellent approximation to the
long-term ($t>> t_0$) form of $\tilde{\phi}(p,t)$ can be derived by
first considering the linear combination of free-particle Gaussian 
momentum-space solutions, namely
\begin{eqnarray}
\phi(p,t)  & = &   \frac{N}{\sqrt{2}}
\left( 
\sqrt{\frac{\alpha}{\sqrt{\pi}}} \, e^{-\alpha^2 p^2/2}\,e^{-ip^2t/2m\hbar}
\, e^{-ipx_0/\hbar}
-
\sqrt{\frac{\alpha}{\sqrt{\pi}}} \, e^{-\alpha^2 p^2/2}\,e^{-ip^2t/2m\hbar}
\, e^{+ipx_0/\hbar} 
\right)
\nonumber \\
& = & 
-2Ni\sqrt{\frac{\alpha}{\sqrt{2\pi}}}
\, 
\sin\left(\frac{p x_0}{\hbar} \right)
\, 
e^{-\alpha^2 p^2/2}\,e^{-ip^2t/2m\hbar}
\, .
\label{long_term_momentum}
\end{eqnarray}
The two contributions differ only in their central location, determined
by the $\exp(\pm i px_0/\hbar)$ terms, corresponding to the $\psi(x,t)$ and
$\psi(-x,t)$ terms. The normalization factor, $N$,  is 
exponentially close to unity if $2x_0 >> \beta$ and 
$\alpha \equiv  \beta/\hbar$. For the `mirror' solution 
in Eqn.~(\ref{original_mirror_solution}), the Fourier transform 
of $\tilde{\psi}(x,t)$ required
to obtain $\tilde{\phi}(p,t)$ initially only samples the isolated 
Gaussian term peaked at $+x_0$, giving the standard Gaussian form for both
$\tilde{\psi}(x,0)$ and $\tilde{\phi}(p,0)$ as shown in 
Fig.~\ref{fig:zero_momentum}. For very long times, however, when both terms
in the `mirror solution' have had time to expand and overlap 
substantially, the Fourier integration over $x \geq 0$ increasingly samples 
contributions
from both $\psi(x,t)$ and $\psi(-x,t)$ more evenly, giving a result
for the associated $\tilde{\phi}(p,t)$ which is close to that in
Eqn.~(\ref{long_term_momentum}), but with an additional factor of $\sqrt{2}$
(to ensure proper normalization) 
since it's eventually restricted to $p>0$.
This gives an approximate expression for the momentum-space probability 
density of
\begin{equation}
|\tilde{\phi}(p,t>>t_0)|^2 \approx \frac{4\alpha}{\sqrt{\pi}}
\,\sin\left(\frac{p x_0}{\hbar}\right)^2\, e^{-\alpha^2 p^2}
\label{long_term_solution}
\end{equation}
for $p>0$ and zero otherwise, and this form exhibits zeros at integral 
multiples of $p = \pi \hbar /x_0$. This limiting case
is also plotted in Fig.~\ref{fig:zero_momentum} (as the dashed curves on the
right) and the numerically obtained results do approach that form for long
times. The long-term value of the expectation value of momentum can also 
be approximated by noting that
\begin{equation}
\langle p \rangle_{t>>t_0}
\approx
\int_{0}^{+\infty}
p \,
\left[
\frac{4\alpha}{\sqrt{\pi}}
\,
\sin\left(\frac{px_0}{\hbar}\right)^2\, e^{-\alpha^2 p^2}
\right]
\, dp
\approx 
\frac{1}{\alpha \sqrt{\pi}}
\end{equation}
where we approximate the oscillating $\sin^2(px_0/\hbar)$ term by its
average of $1/2$. This then gives 
\begin{equation}
\Delta p_{t>>t_0} 
\approx 
\sqrt{\frac{(\pi-2)}{2\pi\alpha^2}}
= 
\left[
\sqrt{\frac{\pi-2}{\pi}}
\right]
\left(
\frac{\hbar}{\sqrt{2}\beta}
\right)
\approx 0.6 \Delta p_0
\end{equation}
which is what is observed numerically for $t>>t_0$ and which is close
to the crude estimate mentioned above. (Note that a similar
expression was found in Ref.~\cite{doncheski_1} for the spread in 
position of a Gaussian wave packet as it hits an infinite wall 
and is temporarily compressed.)

\newpage

\newpage

\newpage

\newpage


\begin{thebibliography}{99}
%
%
\bibitem{saxon}
Saxon D S 
1968
{\it Elementary Quantum Mechanics} 
(San Francisco: Holden-Day) pp~144-147
%
%
\bibitem{andrews}
Andrews M
1998
{\it Wave packets bouncing off walls} 
Am. J. Phys. {\bf 66} 252-254
%
%
\bibitem{doncheski_1}
Doncheski M A  
and 
Robinett R W
1999
{\it Anatomy of a quantum `bounce'}
Eur. J. Phys. {\bf 20} 29-37
%
%
\bibitem{doncheski_2}
Belloni M,
Doncheski M A,
and 
Robinett R W
2005
{\it Exact results for `bouncing' Gaussian wave packets}
Phys. Scripta {\bf 71} 136-140
%
%
\bibitem{dodonov}
Dodonov V V 
and 
Andreata, M A
2000
{\it Deflection of quantum particles by impenetrable boundary}
Phys. Lett. {\bf A275} 173-181;
2002
{\it Quantum deflection of ultracold atoms from mirrors}
Laser Phys. {\bf 12} 57-70;
Andreata M A 
and 
Dodonov V V
2002
{\it The reflection of narrow slow quantum packets from mirrors}
J. Phys. A. Math. Gen. {\bf 35} 8373-8392 
%
%
\bibitem{kleber}
Kleber M 
1994
{\it Exact solutions for time-dependent phenomena in quantum mechanics}
Phys. Rep. {\bf 236}, 331-393.
%
%
\bibitem{born}
Born  M
1955
{\it Continuity, determinism, and reality},
Kgl. Danske Videns. Sels. Mat.-fys. Medd., {\bf 30}~(2) 1;
Born was addressing concerns made by 
A. Einstein 1953 
{\it Elementare \"{U}berlegungen zur Interpretation der Grundlagen der
Quanten-Mechanik}, in {\it Scientific papers presented to Max Born} 
(Edinburgh: Oliver and Boyd) pp~33-40;
Born M 
and 
Ludwig W
1958
{\it Zur Quantenmechanik der kr\"{a}ftefreien Teilchens},
Z. Phys. {\bf 150} 106-117
%
%
\bibitem{equilateral} 
Mathews J 
and 
Walker R L
1970
{\it Mathematical Methods of Physics}, 2nd edition 
(Menlo Park: W. A. Benjamin) pp.~237-239;
Jung C
1980
{\it An exactly soluble three-body problem in one-dimension} 
Can. J. Phys. {\bf 58} 719-728;
Richens P J 
and 
Berry, M V
1981
{\it Pseudointegrable systems in classical and quantum mechanics},
Physica {\bf 2D} 495-512;
Li W -K and Blinder S M
1987
{\it Particle in an equilateral triangle: 
Exact solution of a nonseparable problem},
J. Chem. Educ. {\bf 64} 130-132;
Doncheski M A
and 
Robinett R W
2002
{\it Quantum mechanical analysis of the equilateral triangle billiard:
periodic orbit theory and wave packet revivals}, 
Ann. Phys. (New York) {\bf 299} 208-227 
[arXiv:quant-ph/0307063]
%
%
\bibitem{blueprint}
Doncheski M A,
Heppelmann S,
Robinett R W,
and
Tussey D C
2003
{\it Wave packet construction in two-dimensional quantum billiards:
Blueprints for the square, equilateral triangle and circular cases}
Am. J. Phys. {\bf 71}, 541-557 [arXiv:quant-ph/0307070]
%
%
\bibitem{original_bec} 
Andrews M R,
Townsend C G, 
Miesner H -J,
Durfee D S, 
Kurn D M, 
and 
Ketterle W
1997
{\it Observation of interference between two Bose condensates}
Science {\bf 275} 637-641;
Durfee D S 
and 
Ketterle W
1998
{\it Experimental studies of Bose-Einstein condensates}
Optics Express {\bf 2} 299-313
%
%
\bibitem{other_bec}
Hagley E W,
Deng L,
Kozuma M,
Trippenbach M,
Band Y B,
Edwards M,
Doery M,
Julienne P S,
Helmeson K,
Rolston S L
and 
Phillips W D
{\it Measurement of the coherence of a Bose-Einstein condensate}
1999
Phys. Rev. Lett. {\bf 83} 3112-3115
%
%
\bibitem{bec_many}
Hadzibabic Z,
Stock S, 
Battelier B, 
Bretin, V, 
and 
Dalibard J
{\it Interference of an array of Bose-Einstein condensates}
2004
Phys. Rev. Lett. {\bf 93} 180403
%
%
\bibitem{wallis}
Wallis H, 
R\"{o}hrl A, 
Naraschewski M, and 
Schenzle A
1997
{\it Phase-space dynamics of Bose condensates: Interference versus
interaction}
Phys. Rev. {\bf A55} 2109-2119
%
%
\bibitem{robinett_bec}
Robinett R W
2005
{\it Self-interference of a single Bose-Einstein condensate
due to boundary effects}
to appear in Physica Scripta [arXiv:quant-ph/0511075]
%
%
\bibitem{momentum_space}
Pitaevskii L
and
Strigari S
1999
{\it Interference of Bose-Einstein condensates in momentum space}
Phys. Rev. Lett. {\bf 83} 4237-4240
%
%
\end{thebibliography}
\end{document}